\documentclass{ws-mpla}
\usepackage{graphicx}
\usepackage[usenames ,dvipsnames]{xcolor}
\begin{document}

\markboth{Geng, Huang and Tsai}
{Cosmic Ray Excesses from Multi-component Dark Matter Decays}

%%%%%%%%%%%%%%%%%%%%% Publisher's Area please ignore %%%%%%%%%%%%%%
\catchline{}{}{}{}{}
%%%%%%%%%%%%%%%%%%%%%%%%%%%%%%%%%%%%%%%%%%%%%%%%%%%%%%%%%%%%%%%%%%%

\title{Cosmic Ray Excesses from Multi-component Dark Matter Decays
}

\author{Chao-Qiang~Geng, Da Huang and Lu-Hsing Tsai}
%\footnotesize FIRST AUTHOR\footnote{
%Typeset names in 8 pt Times Roman, uppercase. Use the footnote to
%indicate the present or permanent address of the author.}}

\address{
College of Mathematics \& Physics, Chongqing University of Posts \& Telecommunications, Chongqing, 400065 China\\
Department of Physics, National Tsing Hua University, Hsinchu, 300 Taiwan\\
 Physics Division, National Center for Theoretical Sciences, Hsinchu, 300 Taiwan
}
%author@emailaddress}

%\author{SECOND AUTHOR}
%\address{Group, Laboratory, Address\\
%City, State ZIP/Zone, Country}

\maketitle

%\pub{Received (Day Month Year)}{Revised (Day Month Year)}

\begin{abstract}
We use multi-component decaying dark matter (DM) scenario to explain the possible cosmic ray
excesses in the positron fraction recently confirmed by AMS-02 and the total $e^+ +e^-$ flux observed by Fermi-LAT.
In the two-component DM models, we find an interesting variation of the flavor structure along with the cutoff of the heavy DM.
For the three-component DM case, we focus on a particular parameter range in which the best fits prefer to open only 2 DM decay channels
with a third DM contributing nothing to the electron and positron spectra. We show that all models give the reasonable fits
 to both the AMS-02 positron fraction and the Fermi-LAT total $e^++e^-$ flux, which are also consistent with the measured
 diffuse $\gamma$-ray flux by Fermi-LAT.

\end{abstract}

\ccode{PACS Nos.: 95.35.+d, 98.70.Sa, 13.85.Tp, 14.80.-j}

\section{Introduction}
Despite an overwhelming evidence for the existence of dark matter (DM) in our universe~\cite{Bertone:2004pz},
%most of which comes from the gravitational interactions, while
its real nature is still a great mystery as it has been only seen through gravitational interactions.
%in the universe.
 From the particle physics point of view, it is generally expected that DM is composed of some stable particles
 or the ones with the lifetime much longer than the age of the universe. Since these particles are distributed in our galaxy
 and the universe, it is widely believed that their annihilations and decays would give rise to the visible signals in terms of light stable particles,
 such as positrons/electrons, (anti)protons, photons and neutrinos, which can be observed on the Earth. As a result, such an indirect search
 for DM~\cite{Cirelli:2012tf} is regarded as one of the most promising ways to detect DM.
 % and its particle nature.

Recently, there has been a great advance in the DM indirect detections. For example, the excesses of positrons/electrons were
 discovered by many experiments, such as AMS-01~\cite{AMS01}, ATIC~\cite{ATIC}, PAMELA~\cite{PAMELA,PAMELA2}, Fermi-LAT~\cite{FermiLAT,FermiLAT1,FermiLATp}, and so on, and further confirmed last year by the AMS-02 collaboration~\cite{AMS02} with the unprecedented precision.
 These excesses can be interpreted as the possible signals as DM annihilations~\cite{DMindependent,annihilation,annihilationAMS,AnnihilationDecay}
  or decays~\cite{decay,Ishiwata:2009vx,decayAMS,3bodydecay,3bodydecayAMS,Chen:2009gd,2body}
   although the astrophysical sources like pulsars~\cite{pulsar} would also provide the solution.

The previous studies on the DM annihilating/decaying explanations to the $e^+/e^-$ excesses showed many constraints and limitations. One of the most stringent constraints comes from the measurement of the antiproton flux and the $\bar{p}/p$ ratio by PAMELA~\cite{PAMELApr}, which agrees with the conventional astrophysical theory very well. It clearly implies that DM annihilation/decay should not disturb the (anti)proton flux spectrum much. One simple way to achieve this is the so-called leptophilic DM framework which only allows DM to couple to the lepton sector of the Standard Model (SM), rather than to quarks and gauge bosons. However, even in this framework, the simplest scenario in which a single DM component annihilating or decaying into lepton pairs cannot fit the precise spectra of the Fermi-LAT total $e^+ + e^-$ flux and the AMS-02 positron fraction simultaneously~\cite{Feng:2013zca,2bodyAMSa,2bodyAMSb}. In the literature, several scenarios
have been proposed in order to reduce this tension, such as those with three/four-body decaying/annihilating~\cite{3bodydecayAMS} or asymmetric decaying channels~\cite{asymmetricDM}, and dynamical DMs~\cite{dynamicalDM} as well as other astrophysical solutions~\cite{pulsar,Feng:2013zca}.

In Ref.~\cite{2DM_1}, we have given a multi-component DM scenario which can overcome the difficulty of accommodating the AMS-02 positron fraction with the Fermi-LAT total flux data, while other aspects of multi-component DM models have been investigated in 
Refs.~\cite{dynamicalDM,multiDM,DoubleDiskDM}. In particular, we have shown that two DM components with the heavy DM decaying solely via the $\mu$-channel and the light one predominantly via the $\tau$-channel with the energy cutoff at $E_{cL}=100$ GeV could already give a reasonable fit to both spectra of the AMS-02 positron fraction and the Fermi-LAT total $e^++e^-$ flux. Another advantage for this two-component DM model is that the apparent substructure at around 100 GeV in the above two spectra can have the simple explanation that the light DM drops at that energy. We have also demonstrated that the predicted diffuse $\gamma$-ray spectrum with the best-fit parameters in the $e^+/e^-$ spectrum does not exceed the recent Fermi-LAT bound~\cite{FermiLAT_Gamma}, and is well consistent with the measured one by Fermi-LAT.

In this report, we extend our discussions on the multi-component DM scenario in Ref.~\cite{2DM_1}.
Note that in our previous treatment of the two-DM case, we have fixed our heavy DM energy cutoff to be 1500 GeV and only opened the $\mu$-channel with it.
Although the choice of the parameter and channel is already enough to provide a good fit, it is not quite generic. It is much better to relax such conditions by allowing the energy cutoff to vary and the three leptonic channels to be active in the fitting procedure. One interesting result of this analysis is the change of the best-fitting flavor structure with the different values of the heavy DM cutoff. We will demonstrate that the best-fitting parameters would give $\chi^2/\rm{d.o.f.} \simeq 1$ for the combined AMS-02/Fermi-LAT data, while the predicted diffuse $\gamma-$ray spectra agree with the Fermi-LAT measurement well. The three-component DM case is also explored in a specific example, in which on the basis of the previous two-DM case with the energy cutoffs 100~GeV and 1500~GeV, a third DM is added with the intermediate cutoff lying in the range between 500~GeV and 900~GeV. We find that this additional DM contributes nothing to the final $e^+/e^-$ flux spectra in the best-fit point favored by the combined AMS-02/Fermi-LAT data.

This paper is organized as follows. In Sec. 2, we introduce and analyze our multi-component decaying DM models. In Sec. 3, we discuss the diffuse $\gamma$-ray flux spectra and compare our results with the Fermi-LAT measurement. We give the conclusions
%and some further discussions
in Sec. 4.

\section{AMS-02/Fermi-LAT $e^+/e^-$ Signals from Multi-Component Dark Matter Decays}
\subsection{Framework}
%General Discussions}
Let us begin our discussions by reviewing the computation of the fluxes of electrons/positrons and diffuse $\gamma$-rays in our multi-component DM
framework. In general, the total electron flux $\Phi^{\rm{tot}}_e$ is composed of primary, secondary and DM-decay-induced electrons, while for the positrons in the cosmic rays (CRs), only secondary positrons and the ones from the DM decays contribute, which can be parametrized as follows:
\begin{eqnarray}
\Phi^{(\rm{tot})}_{e} &=& \kappa \Phi^{(\rm{primary})}_e + \Phi^{(\rm{secondary})}+\Phi^{\rm{DM}}_e , \nonumber\\
\Phi^{(\rm{tot})}_p &=& \Phi^{(\rm{secondary})}_p + \Phi^{\rm{DM}}_p.
\end{eqnarray}
It is widely believed that the primary electrons come directly from the supernova remnants distributed in our Galaxy~\cite{SNR}, and their injection spectrum is usually assumed to be a broken power-law function with respect to the rigidity. Note that in order to take into account the normalization uncertainty in the primary electrons, we insert a parameter $\kappa$ which would be determined in the following fitting procedure. Secondary electron/positron fluxes $\Phi^{(\rm{secondary})}_{e,p}$ are the final products of the collisions of the charged particles in the CRs, such as protons and other nuclei, with the interstellar medium (ISM) in the Galaxy. In the present work, we use the GALPROP code~\cite{GALPROP} to simulate the productions and propagations of these background electrons and positrons. For the details of the calculations, especially the choice of the astrophysical parameters, we refer to our earlier work in Ref.~\cite{2DM_1}. The borderline of the calculations is that the spectra of both the positron fraction and the total $e^+ + e^-$ flux are the decreasing power-law functions with respect to the energies.

For the DM signals $\Phi^{\rm{DM}}_{e,p}$, we assume that all of the DM components dominantly decay via the two-body leptonic processes $\chi_i \rightarrow l^{\pm} Y^{\mp}$, where $\chi_i$ denotes the $i$-th DM component, $l = e,\,\mu, \, \tau $, and $Y$ is a charged particle with its mass taken to be $M_Y=300$~GeV. %Such DM decay processes are illustrated in Fig.~\ref{Fig_DMDecay}.
Now, we can express the DM-decay induced source terms $Q^{\rm{DM}}_{e,p}$ as:
\begin{eqnarray}\label{source}
Q({\bf x},p)^{\rm{DM}}_{e,p} = \sum_i \frac{\rho_i({\bf x})}{\tau_i M_i} \Big( \frac{d N_{e,p}}{dE} \Big),
\end{eqnarray}
where $M_i$, $\tau_i$ and $\rho_i({\bf x})$ denote the mass, lifetime and energy density distribution for the $i$-th DM component, respectively. For simplicity, we assume that for the N-component DM models, each DM would carry the same fraction of the total energy density $\rho({\bf x})$ in our Galaxy, implying that $\rho_i({\bf x}) = \rho({\bf x})/N$. Moreover, we shall always take $\rho({\bf x})$ to be of the isothermal profile~\cite{isothermal}. In Eq.~\ref{source}, $(dN_{e,p}/dE)_i$ is the differential $e^-/e^+$ multiplicity per annihilation, which can be expressed as the linear combination of the aforementioned three leptonic channels:
\begin{eqnarray}\label{Norm_Spectrum}
\Big({dN_{e,p}\over dE}\Big)_i={1\over2}\Big[\epsilon^e_i\Big({dN^e\over dE}\Big)_i+\epsilon^\mu_i\Big({dN^\mu\over dE}\Big)_i+\epsilon^\tau_i\Big({dN^\tau\over dE}\Big)_i\Big]\;,
\end{eqnarray}
where $\epsilon^{e,\mu,\tau}$ represent the corresponding branching ratios with the normalization condition $\epsilon^e_i + \epsilon^\mu_i + \epsilon^\tau_i =1$ and the factor $1/2$ accounts for the fact that $e^+$ and $e^-$ arise from two different channels. Since the processes in our discussion are all two-body decays, the final spectrum can be determined solely by the kinematics, irrelevant to any other details of the underlying theory. For $e$- and $\mu$-channels, the normalized $e^\pm$ energy spectra $dN^{e,\mu}/dE$ can be expressed analytically as follows:
\begin{eqnarray}
\Big( \frac{dN^e}{dE} \Big)_i &=& \frac{1}{E_{ci}}\delta(1-x),\\
\Big( \frac{dN^\mu}{dE} \Big)_i &=& \frac{1}{E_{ci}}[3(1-x^2)-\frac{4}{3}(1-x)]\theta(1-x),
\end{eqnarray}
with $x=E/E_{ci}$, while $dN^\tau/dE$ can only be obtained by the simulation of the $\tau$ decay with PYTHIA~\cite{PYTHIA}. After being generated from the DM decays, the positrons and electrons will propagate through a long way in our Galaxy until they are detected by our satellites and telescopes. The propagations of $e^+/e^-$ are very complex~\cite{DiffuseEquation}, during which they would be deflected by the galaxy magnetic fields and lose their energies via the inverse Compton (IC) scattering, bremsstrahlung and synchrotron radiation. In the present work, these complicated propagations are solved with the GALPROP codes by dealing with all of these effects in a consistent way. Note that in our practical calculation we use the same parameter set for the $e^+/e^-$ diffuse propagations as those listed in Table I in Ref.~\cite{2DM_1}. Finally, it is generally expected that the positron and electron fluxes are suffered from the solar modulation, especially for those with energies below 10 GeV. Here, we apply the simple force-field approximation~\cite{SolarModulation} to account for such effects with the potential $\phi_F=0.55$~GV.

\subsection{AMS-02/Fermi-LAT with Two-Component Dark Matter}\label{2DMcase}
As already shown in Ref.~\cite{2DM_1}, two DM components, denoted by ${\rm DM}_{L(H)}$, representing the light (heavy) DM, are enough to accommodate the AMS-02 positron fraction and the Fermi-LAT total $e^+ + e^-$ flux simultaneously. Note that in that analysis, the energy cutoffs of the two DM components were fixed to be 100~GeV and 1500~GeV, respectively, and for the heavy DM component only $\mu$-channel was opened, in order to simplify the calculation. Obviously, such a analysis is not very generic and it is more appropriate to investigate more possibilities. This direction is pursued in the present study.

Here, we still fix the cutoff $E_{cL}$ of the light component ${\rm DM}_{L}$ to be 100~GeV, which has the advantage that the substructure around 100~GeV shown in both the AMS-02 and Fermi-LAT data could be explained as the contribution to $e^+/e^-$ from the light DM terminating exactly there. However, for the cutoff $E_{cH}$ of the heavy component ${\rm DM}_H$, it is taking to be free in the energy range from 1000~GeV to 2000~GeV since the Fermi-LAT total $e^+ + e^-$ flux is extended to as high as 1000~GeV. Furthermore, we let the entire 3 leptonic decay channels active, and pay a special attention to the variation of the flavor structure for the best-fitting points for each $E_{cH}$.

\begin{figure}
\centering
\includegraphics[width=0.33\textwidth, angle =-90]{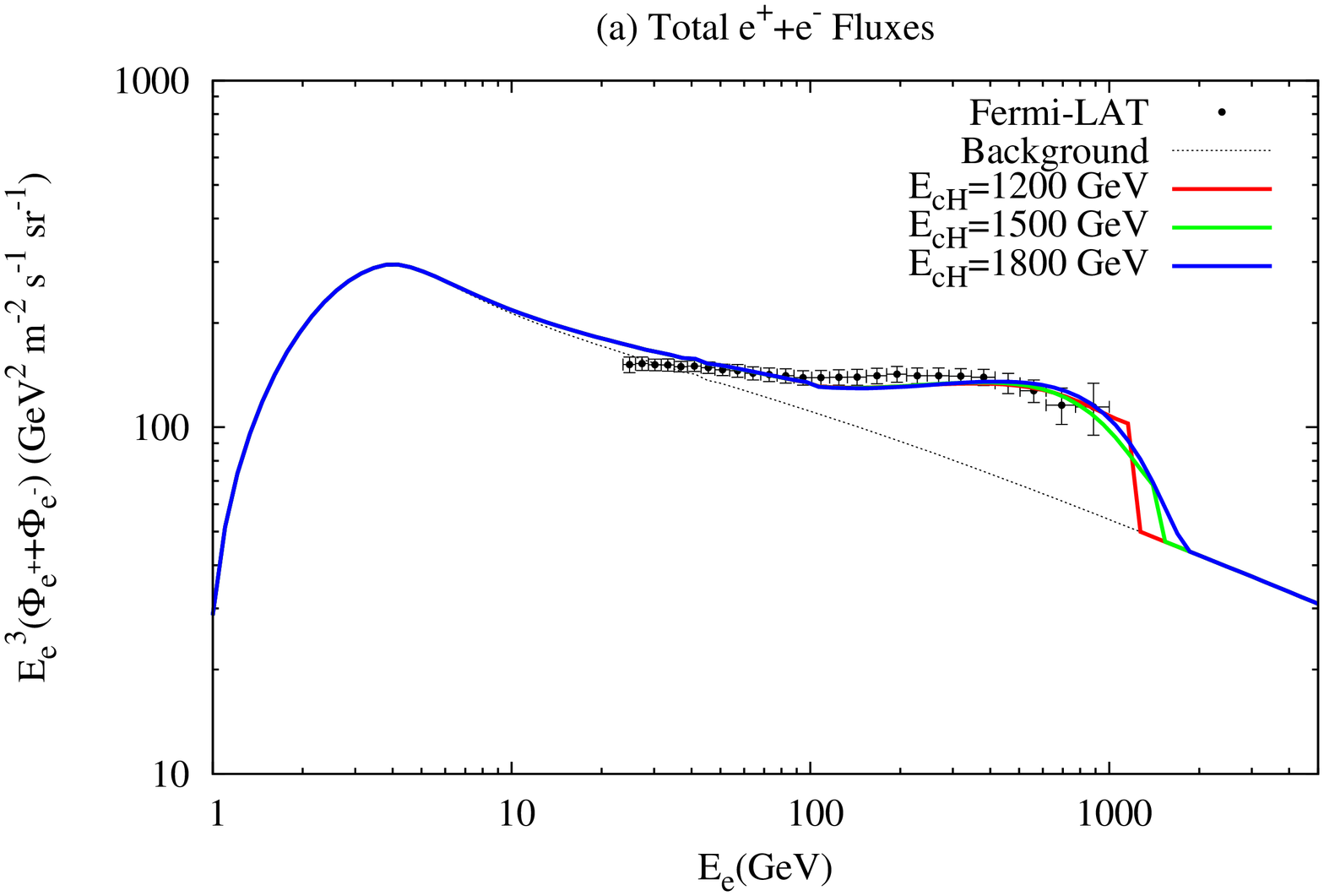}
\includegraphics[width=0.33\textwidth, angle =-90]{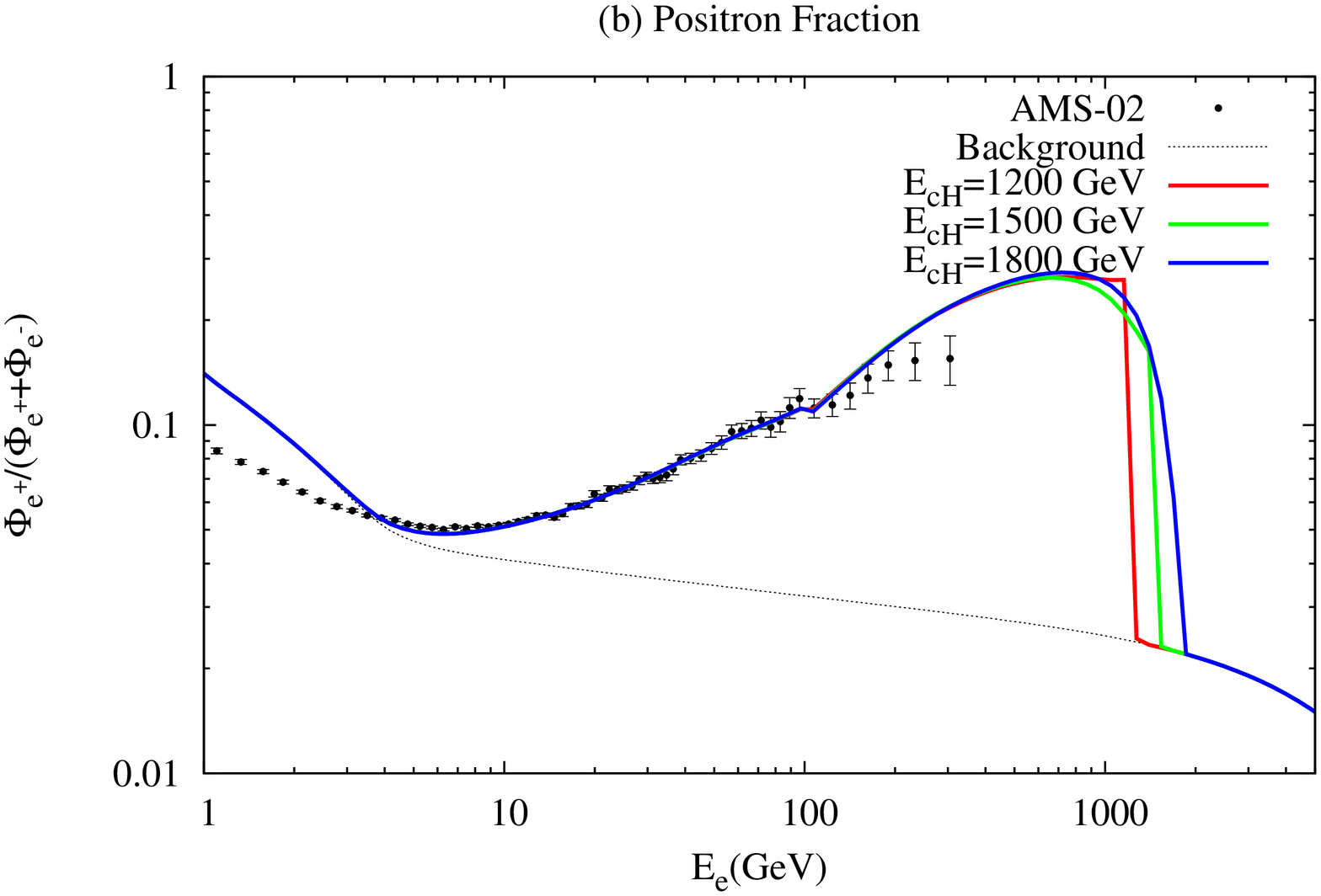}
\caption{(a) Total $e^+ + e^-$ flux and (b) positron fraction from the two-component DM contributions with the best-fitting parameters given in Table~\ref{tab_2DM} for $E_{cH}=$1200 GeV, 1500 GeV and 1800 GeV, respectively.
%, {\color{red} where the black and red dots in (a) represent the  total flux data from Fermi-LAT and  AMS-02, respectively}
}\label{Fig_2DM}
\end{figure}

For the fitting procedure, we apply the simple $\chi^2$-minimization method, and take the 26 data points of the total electron/positron flux from Fermi-LAT and the 42 data points of the positron fraction from AMS-02 with the energy above 10~GeV to reduce the influence of the solar modulation.
In Table~\ref{tab_2DM} and Fig.~\ref{Fig_2DM},
we present the best-fit results with $\chi_{\rm min}^2$=62.4, 62.3 and 62.8 and $\chi_{\rm min}^2/d.o.f.$ all around 1.06
for the three typical examples with the cutoffs and the masses of the heavy DM at ($E_{cH}$,$M_H$)=(1200, 2436) GeV, (1500,3030) GeV and (1800,3624) GeV, respectively.
\begin{table}[h]
\tbl{Parameters leading to the minimal values of $\chi^2$ with the cutoffs of heavy DM being 1200, 1500 and 1800 GeV, respectively.}
%The point in the parameter space which gives the minimal value of $\chi^2$ with the cutoffs of heavy DM taken as 1200, 1500, 1800 GeV, respectively }
{\begin{tabular}{@{}cccccc@{}} \toprule
%[t]{l|c|cccc}
%\hline
$E_{cH}({\rm GeV})$& $\kappa$ & $\epsilon^{e}_{H,L}$& $\epsilon^{\mu}_{H,L}$ & $\epsilon^{\tau}_{H,L}$
 & $\tau_{H,L}(10^{26}{\rm s})$
 %& $\epsilon^{e}_L$& $\epsilon^{\mu}_L$ & $\epsilon^{\tau}_L$ &$\tau_L(10^{26}{\rm s})$
%& $\chi_{\rm min}^2$& $\chi_{\rm min}^2/d.o.f.$\\
\\
\colrule
%\hline
1200 & 0.844 & 0.206,0.015 & 0.794,0 & 0,0.985 & 0.97,0.83\\
% & 0.015  & 0 &  0.985 & {0.83} \\
%& 62.4& 1.06\\
1500 & 0.844 & 0.058,0.020 & 0.942,0 & 0,0.980 & 0.78,0.82\\
% & 0.020  & 0 &  0.980 & {0.82} \\
% & 62.3& 1.06\\
1800 & 0.843 & 0,0.022 & 0.842,0 & 0.158,0.978 & 0.64,0.83\\
% & 0.022  & 0 &  0.978 & {0.83} \\
% & 62.8& 1.06\\
%\hline
\botrule
\end{tabular}\label{tab_2DM}}
\end{table}
%From Table~\ref{tab_2DM}, we can see that
Note that all three cases essentially give the same degree of the goodness of the fitting,
 %with the minimum $\chi_{\rm min}^2/d.o.f.$ around 1.06,
 indicating that the fitting results are very reasonable.
However, what is interesting here is the modification of the flavor structure with the different choices of the cutoff for
the heavy DM component. Specifically, when the heavy DM cutoff $E_{cH}$ is chosen to be smaller than 1500 GeV, quite a considerable portion of $e^+/e^-$ comes from the heavy DM decay via the electron channel. And when $E_{cH}$ becomes larger, the heavy DM contribution to $e^+/e^-$
from the $e$-channel would decrease, finally terminating at around 1500 GeV, whereas the positrons/electrons from the $\tau-$channel would arise and increase. This feature shows the power of the precise measurement in the indirect DM search experiments, from which we can obtain a lot of information of the DM properties.

\subsection{AMS-02/Fermi-LAT with Three-Component Dark Matters}\label{3DMcase}
In general, it is expected that new DM components would result in better fittings for the $e^+/e^-$ spectra due to
the more free parameters
%, resulting in a better fitting, as the $e^+/e^-$ spectra are
from the mixture of those of DM components. However, we shall illustrate an interesting counterexample in this paper. If we fix two energy cutoffs of the three DMs to be 100 GeV and 1500 GeV as before and add a middle-mass DM with its energy cutoff lying in the range between 500 GeV and 900 GeV,
 the minimum $\chi^2$ with the AMS-02/Fermi-LAT data yields that the middle-mass DM does not contribute any electrons and positrons to the final $e^+/e^-$ fluxes. Table~\ref{tab_3DM} shows a typical example of this case, in which the cutoff and the mass of the extra DM are fixed to be ($E_{cM}$, $M_M$)=(760, 1577)~GeV.
\begin{table}[h]
\tbl{Parameters which give the minimal values of $\chi^2$ with the cutoffs and masses of the three DMs: ($E_{cL}$, $M_L$)=(100, 416)~GeV, ($E_{cM}$, $M_M$)=(760, 1577)~GeV and ($E_{cH}$, $M_H$)=(1500, 3030)~GeV, respectively.}
%The point in the parameter space which gives the minimal value of $\chi^2$ with the cutoffs and masses of the three DMs taken to be ($E_{cL}$, $M_L$)=(100, 416)~GeV, ($E_{cM}$, $M_M$)=(760, 1577)~GeV and ($E_{cH}$, $M_H$)=(1500, 3030)~GeV, respectively }
{\begin{tabular}{@{}ccccccc@{}} \toprule
%{\begin{tabular}[t]{l|cccc|cccc|cccc|cc}
%\hline
 $\kappa$ & $\epsilon^{e}_{H,M,L}$& $\epsilon^{\mu}_{H,M,L}$ & $\epsilon^{\tau}_{H,M,L}$
 & $\tau_{H,M,L}(10^{26}{\rm s})$
% & $\epsilon^{e}_M$& $\epsilon^{\mu}_M$  & $\epsilon^{\tau}_M$
% & $\tau_M(10^{26}{\rm s})$ & $\epsilon^{e}_L$ & $\epsilon^{\mu}_L$ & $\epsilon^{\tau}_L$
%&$\tau_L(10^{26}{\rm s})$
& $\chi_{\rm min}^2$& $\chi_{\rm min}^2/d.o.f.$\\
\colrule
0.844 & 0.058,0,0.020 & 0.942,0,0 & 0,0,0.980 & 0.52,$\infty$,0.55 & 62.3& 1.06\\
\botrule
\end{tabular}\label{tab_3DM}}
\end{table}
It is seen that the best-fitting values prefer the three leptonic channels of the middle-mass DM to be all closed. The only difference is the lifetimes of two remaining DMs, which are just a factor 2/3 of those in the corresponding two-DM case in order to compensate the less energy density for each DM in Eq.~(\ref{source}). Thus, the resultant spectra of the positron fraction and the $e^+ + e^-$ flux are exactly the same as the corresponding two-DM case with ($E_{cL}$, $M_L$)=(100, 416)~GeV and ($E_{cH}$, $M_H$)=(1500, 3030)~GeV. The above discussion implies that in this specific range of the three-DM parameter space, the corresponding two-DM models are more statistical favored.

\section{$\gamma$-Ray Fluxes From Multi-Component Dark Matter}
The positrons and electrons from DM decays or annihilations are inevitable to be followed with the emissions of high energy photons manifested in the measured diffuse $\gamma$-ray spectrum~\cite{Ishiwata:2009vx,Cirelli:2012ut,Papucci:2009gd,Beacom:2004pe,Gamma_DM,Gamma_extraDMIC,Gamma_extra,Ibarra:2007wg}. However, the produced $\gamma$-ray flux should not exceed the bounds obtained by the diffuse $\gamma$-ray data by Fermi-LAT~\cite{FermiLAT_Gamma} and EGRET~\cite{EGRET}. Thus, as pointed in Refs.~\cite{Cirelli:2012ut,Papucci:2009gd,Gamma_DM,Gamma_extraDMIC,Gamma_extra}, these diffuse $\gamma$-ray constraints have already excluded a large portion of the parameter space of the decaying DM models to explain the PAMELA/Fermi-LAT $e^+/e^-$ excesses. Nevertheless, in Ref.~\cite{2DM_1}, we have demonstrated that our multi-component DM scenario survives even under the stringent $\gamma$-ray constraints.
%In this section, we present our numerical results for the three benchmark points listed in Table~\ref{tab_2DM} of the two-component DM models. For the three-component DM case studied in Section~\ref{3DMcase}, since the predicted $\gamma$-ray spectra are the same as the corresponding two-DM models, we do not show them separately.

The diffuse $\gamma$-ray flux in our present model has many components, including the usual astrophysical diffuse background $\gamma$-ray radiation and DM contributions inside and outside our Galaxy. The final spectrum will be compared with the Fermi-LAT inclusive continuum photon spectrum~\cite{FermiLAT_Gamma}.
%, which was measured within the energy range $4.8~\mbox{GeV} < E_\gamma < 264~\mbox{GeV}$ covering the sky in the high latitude with $|b|>10^\circ$ plus the Galactic center (GC) with $|b|<10^\circ$, $l<10^\circ$ and $l>350^\circ$.
The conventional astrophysical background inside the Milky Way includes three sources: pion decay, inverse Compton (IC) scattering and bremsstrahlung. All of them are calculated by the GALPROP code with the same astrophysical parameters already determined when we compute the electron/positron fluxes in the previous section. For the extragalactic $\gamma$-ray background (EGB) originated from the superposition of the unresolved extragalactic sources such as the active galactic nuclei (AGN), we take the form $E^2 d \Phi_\gamma / dE = 5.18\times 10^{-7} E^{-0.499} {\rm GeV \, cm^{-2} \, sr^{-1} \, s^{-1}}$, obtained by fitting the low-energy data from EGRET~\cite{EGRET,Ishiwata:2009vx}.

The DM contributions to the $\gamma$-ray fluxes can also be divided into the parts inside and outside the Galaxy. Inside the Galaxy, the extra electrons and positrons as the decay products of multiple DM components can induce the $\gamma$-rays in the processes such as bremsstrahlung and IC. Moreover, the additional prompt decay processes, such as the final state radiations (FSRs) from the three 
leptons~\cite{Beacom:2004pe,Gamma_DM,FermiLAT_Gamma}, the radiative muon decays~\cite{Gamma_DM} and the pion decays from $\tau$~\cite{Gamma_DM,Fornengo:2004kj}, could give rise to the continuous spectrum of the $\gamma$-rays. The DMs outside the Galaxy are also generally expected to lead to the $\gamma$-rays via the prompt decay processes and the IC scattering with the CMB photons. The only additional effect is the redshift of the spectra caused from the cosmic expansion. For the calculation details, please refer to Ref.~\cite{2DM_1}. Here, we only emphasize that in these calculations of DM diffuse $\gamma$-ray spectra, all of the parameters are already fixed by fitting the AMS-02/Fermi-LAT data, and thus the final spectra are the predictions of our DM models.

\begin{figure}
\centering
\includegraphics[width=0.5\textwidth, angle =-90]{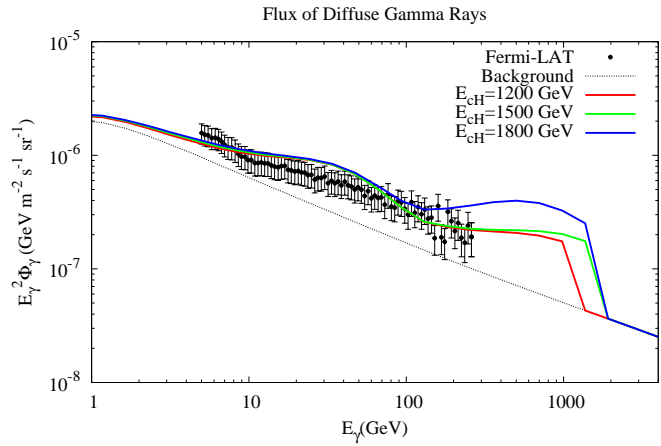}
\caption{Diffuse $\gamma$-ray spectrum from the two-component DM contributions with the best-fitting parameters given in Table~\ref{tab_2DM} for $E_{cH}=$1200 GeV, 1500 GeV and 1800 GeV, respectively.
%, {\color{red} where the black and red dots in (a) represent the  total flux data from Fermi-LAT and  AMS-02, respectively}
}\label{Fig_3DM}
\end{figure}

By summing up the above backgrounds and DM contributions, we can obtain the total diffuse $\gamma$-ray spectra in the multi-component decaying DM models. Fig.~\ref{Fig_3DM} shows the diffuse $\gamma$-ray spectra for the two-component DM models studied in Section~\ref{2DMcase} with the parameters listed in Table~\ref{tab_2DM}. From the table, it is clear that the the predictions of the $\gamma$-ray fluxes in these three cases are consistent with the Fermi-LAT measurement in all the energy range. In order to further support this conclusion, we compute the usual $\chi^2$ with the 82 Fermi-LAT data points. The final results are $\chi^2 =$ 80.4, 83.5 and 151.0 with $\chi^2/{\rm d.o.f.} <2$
for the two-component DM models, in which the cutoffs and masses of the heavy DM are fixed to be ($E_{cH}$, $M_H$) = (1200, 2436)~GeV, (1500, 3030)~GeV and (1800, 3625)~GeV, respectively.
%, all of which have .

\section{Conclusions}
% and Discussions}
We have explored the multi-component decaying DM scenario along the line described in Ref.~\cite{2DM_1}. The extension is in two aspects. First, we allow the energy cutoff of the heavy DM to vary in the range between 1000 and 2000~GeV, followed by the change of the heavy DM mass accordingly, and open all of 6 leptonic decay channels in the two-component DM case. We have found an interesting change of the best-fit heavy DM flavor structure with the different choices of the cutoff and the mass of the heavy DM. With the increase of the heavy DM cutoff in the assigned range, the electron-channel branching ratio of the heavy DM originally with a relatively large size begins to decrease and stop at around 1500~GeV, whereas from around 1500~GeV its $\tau$-channel branching ratio arises and increases until a considerable portion at 2000~GeV. The points in the parameter space all give the reasonable fits to the AMS-02 positron fraction and the Fermi-LAT total $e^+ + e^-$ flux data, with the $\chi^2_{\rm min}/{\rm d.o.f.} \simeq 1$. We have further shown that the predicted diffuse $\gamma$-ray spectra for all these examples are consistent with the Fermi-LAT measurement.

We have also investigated the three-component DM model with the particular focus on a region of the parameter space. Besides the two DMs with the energy cutoffs 100~GeV and 1500~GeV as the two-component case, we have incorporated the a third DM with the cutoff $E_{cM}$ lying in the range between $500$~GeV and $900$~GeV. This parameter space range is interesting because within it, the fittings of other parameters, especially the flavor branching ratios for the three DMs, prefer
the original two-component DM cases. For the best-fitting points, the three leptonic channels related to the middle-mass DM do not contribute any $e^+/e^-$ to the final spectra. This result contradicts our general expectation that the addition of more DMs would lead to a better fitting due to a complicated mixture of the different channels.

It is clear that the present multi-component decaying DM scenario should be further studied. One possible problem is related to the fact that we have only introduced a phenomenological decay coupling of DMs with the effective dimensionless coupling being of {\cal O}($10^{-26}$), which is obviously quite unnatural. In order to examine this problem, we should resort to some more fundamental particle models implemented with the full Lagrangian and symmetries. Another question is about the compatibility of the AMS-02 positron fraction and the Fermi-LAT total $e^++e^-$ flux data. More recently, the AMS-02 collaboration has presented the preliminary measurements on the electron, the positron and the total $e^++e^-$ fluxes. It is more appropriate to use the data sets from a single experiment, say AMS-02. We will study these issues elsewhere.

\section*{Acknowledgments}
The work was supported in part by National Center for Theoretical Science, National Science
Council (NSC-101-2112-M-007-006-MY3) and National Tsing Hua
University (Grant Nos. 102N1087E1 and 102N2725E1).

%\section*{References}

\end{document}